Regular Article - Theoretical Physics

# Observational properties of a Schwarzschild black hole surrounded by a Dehnen-type dark matter halo

Zhi Li[1,2,3,a], Jiancheng Yu[4,b]

[1] Yunnan Observatories, Chinese Academy of Sciences, Kunming 650216, People's Republic of China
[2] Key Laboratory for Structure and Evolution of Celestial Objects, Chinese Academy of Sciences, Kunming 650216, People's Republic of China
[3] International Centre of Supernovae, Yunnan Key Laboratory, Kunming 650216, People's Republic of China
[4] College of Physics, Guizhou University, Guiyang 550025, China

**Abstract** This study investigates the accretion process and observational signatures of thin accretion disks around a Schwarzschild black hole (BH) embedded in a Dehnen-type dark matter (DM) halo. We examine the influence of the density $\rho_s$ and radius $r_s$ of the DM halo on key disk properties, including the energy flux, temperature distribution, and emission spectrum. Our results show that all three of these quantities decrease with increasing $\rho_s$ or $r_s$. Furthermore, by generating and analyzing both direct and secondary images of the accretion disk, we explore how the observational inclination angle and the DM halo parameters $\rho_s$ and $r_s$ affect image profiles. Finally, the observed flux distributions are presented for different inclination angles. Our findings indicate that the accretion disk becomes colder and dimmer as the DM halo parameters increase, highlighting the significant role of DM in shaping BH observables.

## 1 Introduction

The study of BHs lies at the heart of modern theoretical and astrophysical research, drawing interdisciplinary interest. A prevalent scenario in current modeling posits that astrophysical BHs are embedded within extended DM halos. Investigating the interplay between such halos and supermassive BHs is essential for understanding both BH evolution and DM properties. DM halos profoundly shape large-scale dynamics, explaining the flatness of galactic rotation curves [1–4] and providing direct evidence for DM in systems like the Bullet Cluster [5]. On smaller scales, their presence can influence relativistic effects near BHs, including gravitational lensing [6,7], BH shadows, and polarized images [8,9]. Various parameterized models have been proposed to describe the density distribution of dark matter (DM) halos. Among them are the Navarro–Frenk–White (NFW) profile [10], which features a divergent $1/r$ density cusp in the central region, the Einasto profile [11] characterized by a continuously varying logarithmic slope, and the Burkert profile [12], which predicts a constant-density core. For the present study, we adopt the Dehnen-type DM halo model [13–15], which offers a particularly versatile framework due to its three parameters $(\alpha, \beta, \gamma)$. By adjusting the parameters $(\alpha, \beta, \gamma)$, this profile can reproduce a wide range of density behaviors. For instance, it reduces to the Hernquist profile when $(\alpha, \beta, \gamma) = (1, 4, 1)$ and to the Jaffe profile when $(\alpha, \beta, \gamma) = (1, 4, 2)$. This flexibility makes the Dehnen model well-suited for a parametric study aiming to explore the generic influence of a spherically symmetric DM halo on accretion disk properties, without being tied to the specific assumptions of a single fixed profile.

While the gravitational influence of DM halos on galactic-scale phenomena has been extensively studied, their imprint on BH accretion processes remains less explored. Thin accretion disks, where gravitational energy is efficiently converted into observable radiation, serve as powerful probes of the influence of DM halos on BH environments. The foundational model for geometrically thin, optically thick disks was developed by Shakura and Sunyaev [16], who formulated a hydrodynamic framework based on viscous stresses and Keplerian rotation. This was later generalized to a fully relativistic setting by Novikov and Thorne [17,18], yielding the canonical Novikov–Thorne model for steady-state accretion with a constant mass accretion rate. Since then, the energy flux, temperature profile, and emitted spectra of thin disks have been widely investigated in diverse spacetimes, includ-ing those in modified gravity theories and around exotic compact objects [19–34]. These studies underscore the diagnostic power of accretion disks in probing strong-field gravity and constraining new physics [35].

[a] e-mail: lizhi@ynao.ac.cn (corresponding author)
[b] e-mail: gs.jcyu24@gzu.edu.cn

In parallel, significant progress in modeling BH accretion images has been made since the pioneering work of Luminet [36], who computed the direct and secondary images of a thin disk around a Schwarzschild BH using an analytical expression for the radiation flux. Two primary approaches are now employed: semi-analytical methods and numerical ray-tracing combined with radiative transfer. Subsequent advances have led to the development of sophisticated ray-tracing codes incorporating general relativistic effects and radiative transfer, such as those in [37–45]. These tools have enabled detailed studies of thin accretion disks in various spacetimes, including those predicted by modified gravity theories and exotic compact objects [46–56]. Building on these developments, we investigate the impact of Dehnen-type DM halo on the observable properties of thin accretion disks around Schwarzschild BHs, aiming to identify potential signatures of DM in strong-gravity regimes.

To place our findings in a broader context, it is pertinent to compare the effects of the Dehnen-type dark matter halo adopted in this work with other DM models and accretion disk settings explored in the literature. For instance, Boshkayev et al. [57] investigated a Schwarzschild black hole surrounded by a DM halo with tangential pressure-modeled as an Einstein cluster-and found that the accretion disk luminosity is enhanced compared to the vacuum case. This stands in contrast to our results, where the presence of a Dehnen-type halo leads to a systematic suppression of the energy flux, temperature, and spectral luminosity. The difference highlights the significant role played by the DM density profile and the assumed pressure components in shaping the accretion disk's observational signatures. Similarly, studies considering anisotropic DM pressures [58] have shown that the accretion disk's spectral luminosity can be either enhanced or suppressed depending on the sign of the anisotropy parameter, though the dominant effect remains the presence of DM itself. In our Dehnen model, we observe a consistent dimming and cooling of the disk, reinforcing the notion that the gravitational potential of the DM halo is the primary factor influencing the disk's thermodynamic properties, albeit with a net suppressive effect in our specific configuration. Furthermore, comparisons with Hernquist-type DM halos [59] reveal that such profiles also lead to an increase in the ISCO radius-a trend qualitatively consistent with our findings. However, the Hernquist model can lead to an increase in radiative efficiency for small scale radii, whereas our Dehnen model generally results in a cooler and less luminous disk. This divergence underscores the importance of the inner density slope and concentration of the DM halo in determining the accretion dynamics.

The structure of this paper is as follows. In Sect. 2, we review the spacetime geometry of a Schwarzschild BH surrounded by a Dehnen-type DM halo and analyze photon geodesics in the equatorial plane, including the resulting photon orbits. In Sect. 3, the energy flux, temperature distribution, and emission spectrum of thin accretion disks around such a BH are computed within this extended spacetime. In Sect. 4, we employ the $(\varphi(b))$ diagram to examine the imaging of BH accretion disks, plotting both direct and secondary images of the accretion disk and comparing them with those of standard Schwarzschild BHs. Additionally, the apparent radiation flux distribution as seen by distant observers at different inclination angles is examined. In Sect. 5, we summarize our findings.

## 2 A Schwarzschild black hole in a Dehnen-type dark matter halo: geodesic analysis

### 2.1 Schwarzschild black hole surrounded by a Dehnen-type dark matter halo

The Dehnen-type DM halo is characterized by the following density profile:

$$\rho = \rho_s \left(\frac{r}{r_s}\right)^{-\gamma} \left[\left(\frac{r}{r_s}\right)^\alpha + 1\right]^{\frac{\gamma-\beta}{\alpha}}, \qquad (2.1)$$

where $\rho_s$ and $r_s$ represent the central density of the halo and the core radius of the halo, respectively. The parameter $\gamma$ specifies the particular variant of the profile. In the case of Dehnen-$(\alpha, \beta, \gamma) = (1, 4, 5/2)$, the density profile of the Dehnen-type DM halo is

$$\rho = \frac{\rho_s}{\left(\frac{r}{r_s}\right)^{5/2} \left(\frac{r}{r_s} + 1\right)^{3/2}}. \qquad (2.2)$$

The Schwarzschild BH in the Dehnen-(1,4,5/2) DM halo can be written as

$$ds^2 = -f(r)dt^2 + \frac{dr^2}{f(r)} + r^2 \left(d\theta^2 + \sin^2\theta d\phi^2\right), \qquad (2.3)$$

where [60]

$$f(r) = 1 - \frac{2M}{r} - 32\pi \rho_s r_s^3 \sqrt{\frac{r+r_s}{r_s^2 r}}. \qquad (2.4)$$

### 2.2 Analysis of equatorial null and timelike geodesics

The motion of a test particle in curved spacetime is governed by the geodesic equations, which are derived from the Euler-Lagrange equations

$$\frac{d}{d\tau}\left(\frac{\partial \mathcal{L}}{\partial \dot{x}^\mu}\right) = \frac{\partial \mathcal{L}}{\partial x^\mu}, \qquad (2.5)$$

where $\tau$ is the proper time, $\dot{x}^\mu = dx^\mu/d\tau$ is the four-velocity of the test particle, and $\mathcal{L}$ is the Lagrangian. For the spacetime geometry defined by Eq. (2.3), the Lagrangian takes the form

$$\mathcal{L} = \frac{1}{2} g_{\mu\nu} \dot{x}^\mu \dot{x}^\nu. \tag{2.6}$$

Without loss of generality, we restrict our analysis to geodesic motion in the equatorial plane $\theta = \pi/2$, where the spacetime's symmetries yield two conserved quantities: the energy $E$ and the angular momentum $L$. These are given by

$$E = -g_{tt}\dot{t}, \tag{2.7}$$
$$L = g_{\phi\phi}\dot{\phi}. \tag{2.8}$$

The photon geodesic equations are obtained by combining Eqs. (2.6)–(2.8) and introducing the rescaled affine parameter $\tau = \tau/L$. In terms of this parameter, the equations become

$$\dot{t} = \frac{1}{bf(r)}, \tag{2.9}$$
$$\dot{\phi} = \frac{1}{r^2}, \tag{2.10}$$
$$\dot{r}^2 = \frac{1}{b^2} - \frac{f(r)}{r^2}, \tag{2.11}$$

where $b = L/E$ is the impact parameter associated with the light ray. Combining Eqs. (2.10) and (2.11), the photon trajectory in the equatorial plane is governed by

$$\left(\frac{dr}{d\phi}\right)^2 = r^4 \left(\frac{1}{b^2} - \frac{f(r)}{r^2}\right) \equiv V_{\text{eff}}, \tag{2.12}$$

where $V_{\text{eff}}$ is the effective potential of the photon. A light ray's trajectory is fully specified by its impact parameter $b$. The photon sphere radius $r_{ph}$, which arises from a bound orbit of light, is obtained by solving:

$$V_{\text{eff}}|_{r=r_{ph}} = 0, \tag{2.13}$$
$$\frac{dV_{\text{eff}}}{dr}|_{r=r_{ph}} = 0. \tag{2.14}$$

Solving Eqs. (2.13) and (2.14) simultaneously yields the critical impact parameter $b_c$ associated with the photon sphere, given by

$$b_c = \frac{r_{ph}}{\sqrt{f(r_{ph})}}. \tag{2.15}$$

The trajectory of a light ray near a BH is critically determined by its impact parameter $b$ relative to the critical value $b_c$. For $b > b_c$, the photon originates from spatial infinity, approaches a minimum radial distance at the periastron, and is subsequently deflected back to infinity. In contrast, for $b < b_c$, the photon inevitably crosses the event horizon and is captured by the BH. At the precise critical value $b = b_c$, the light ray will surround BHs in the circular orbit.

The study of light bending near a BH is facilitated by introducing the substitution $u = 1/r$. With this transformation, the orbital Eq. (2.12) can be rewritten as

$$\left(\frac{du}{d\phi}\right)^2 = \frac{1}{b^2} - u^2 f\left(\frac{1}{u}\right) \equiv G(u). \tag{2.16}$$

When $b < b_c$, the trajectory plunges into the BH. However, the external part of the orbit enables the determination of the total accumulated azimuthal angle $\varphi$ (The azimuthal coordinate of the spacetime is denoted by $\phi$, and $\varphi$ represents the total azimuthal displacement along the photon's trajectory):

$$\varphi = \int_0^{u_h} \frac{1}{\sqrt{G(u)}} du, \quad b < b_c. \tag{2.17}$$

Here, $u_h = 1/r_h$, where $r_h$ is the radius of the outermost event horizon. For $b > b_c$, the total azimuthal deflection $\varphi$ of a photon trajectory with impact parameter $b$ is given by

$$\varphi = 2\int_0^{u_{min}} \frac{1}{\sqrt{G(u)}} du, \quad b > b_c, \tag{2.18}$$

where $u_{\min}$ is the smallest positive solution to $G(u) = 0$, occurring for impact parameters $b > b_c$.

For photons with impact parameter $b = b_c$, the trajectory asymptotically approaches the photon sphere at $u = u_{ph}$, winding around the BH infinitely many times in the ideal limit. The total azimuthal angle $\varphi$ is plotted as a function of $b$ in Fig. 1. For fixed $\rho_s$, the $\varphi(b)$ curve shifts to larger $b$ as $r_s$ increases. Similarly, increasing $\rho_s$ at fixed $r_s$ also shifts the profile rightward.

Photon trajectories for different parameters are shown in Fig. 2. Trajectories with $\varphi < 1.5\pi$ are plotted in black, those with $1.5\pi < \varphi < 2.5\pi$ in orange, and those with $\varphi > 2.5\pi$ in red, reflecting the degree of orbital winding near the BH.

For timelike geodesics in the equatorial plane ($\mathcal{L} = -1/2$), the orbit equation takes the form

$$\left(\frac{dr}{d\phi}\right)^2 = r^4 \left(\frac{1}{b^2} - \frac{f(r)}{r^2} - \frac{f(r)}{L^2}\right) = \tilde{V}_{\text{eff}}. \tag{2.19}$$

The radius of the innermost stable circular orbit (ISCO) is determined by the effective potential and its derivatives.

$$\tilde{V}_{\text{eff}}|_{r=r_{isco}} = \frac{d\tilde{V}_{\text{eff}}}{dr}|_{r=r_{isco}} = \frac{d^2\tilde{V}_{\text{eff}}}{dr^2}|_{r=r_{isco}} = 0. \tag{2.20}$$

In the analysis of thin accretion disks, the ISCO at $r = r_{isco}$ defines the inner edge of the disk. For $r < r_{isco}$, equatorial circular orbits become radially unstable, and any perturbation leads to a rapid plunge toward the BH. Thus, $r_{isco}$ marks the critical radius beyond which stable orbital motion is no longer possible, establishing the disk's innermost extent in the standard model.

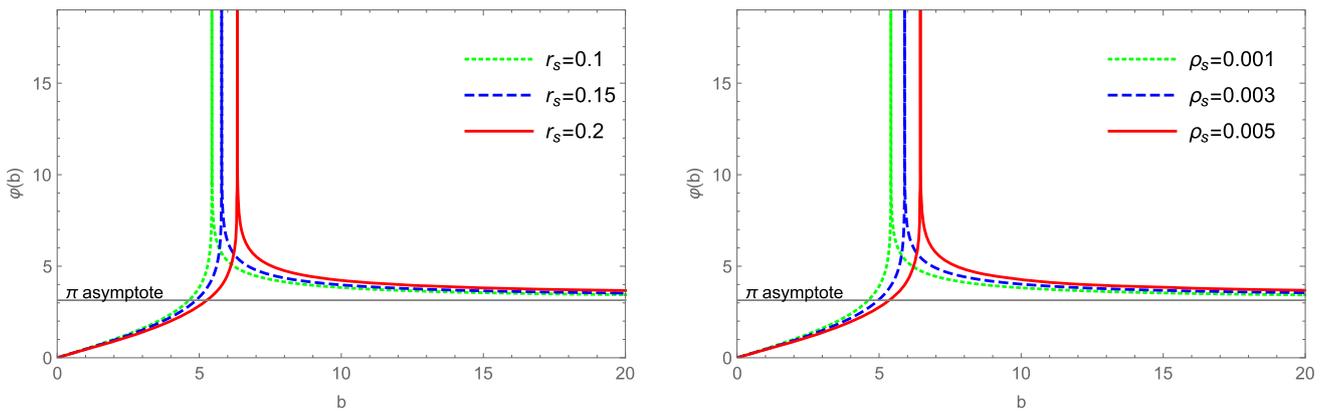

**Fig. 1** The total azimuthal angle $\varphi$ is plotted as a function of the impact parameter $b$ for different values of the parameters. (Left) For fixed $\rho_s = 0.03$ and varying $r_s$. (Right) For fixed $r_s = 0.5$ and varying $\rho_s$

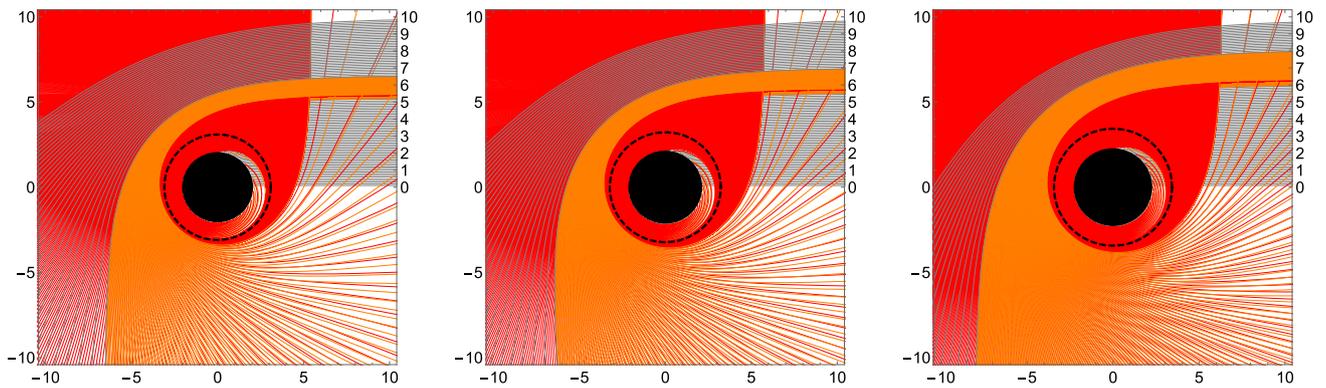

**Fig. 2** Photon trajectories are shown for different values of the parameter $r_s$ at fixed $\rho_s = 0.03$. From left to right: $r_s = 0.1$, 0.15, and 0.2

## 3 The properties of thin accretion disks

In this section, we provide a concise overview of the Novikov–Thorne model, a relativistic extension of the standard Shakura–Sunyaev framework for geometrically thin, optically thick accretion disks. The model adopts the following key assumptions: (1) The central compact object is described by a stationary, axisymmetric, and asymptotically flat spacetime. (2) The accretion disk is treated as a test structure, meaning its self-gravity does not perturb the background metric. (3) The disk is geometrically thin, with a vertical height negligible compared to its radial extent. (4) Matter follows nearly circular geodesics between an inner boundary, defined by the ISCO at $r_{isco}$, and an outer radius $r_{out}$. (5) The disk is aligned with the equatorial plane of the central object. (6) The disk is in local thermodynamic equilibrium, and the radiation emitted from its surface is well approximated by a blackbody spectrum. (7) The mass accretion rate $\dot{M}_0$ remains constant in time, corresponding to a steady-state accretion flow. The physical constants and thin accretion disk parameters are fixed at the following values throughout our computation: $c = 2.997 \times 10^{10}$ cms$^{-1}$, $\dot{M}_0 = 2 \times 10^{-6} M_\odot$ yr$^{-1}$, 1 yr $= 3.156 \times 10^7$ s, $\sigma_{SB} = 5.67 \times 10^{-5}$ ergs$^{-1}$ cm$^{-2}$ K$^{-4}$, $h = 6.625 \times 10^{-27}$ ergs, $k_B = 1.38 \times 10^{-16}$ erg K$^{-1}$, $M_\odot = 1.989 \times 10^{33}$ g, and the mass of BH $M = 2 \times 10^6 M_\odot$. The local energy flux $F(r)$ emitted from the disk surface is calculated using the relativistic accretion disk model of Page and Thorne [17], which provides a self-consistent description of radiative transfer in the Novikov–Thorne framework:

$$F(r) = -\frac{\dot{M}_0 \Omega_{,r}}{4\pi\sqrt{-g}(E - \Omega L)^2} \int_{r_{isco}}^{r} (E - \Omega L) L_{,r} dr, \quad (3.1)$$

where $g$ stands for the metric determinant. $E$, $L$, and $\Omega$ denote the energy, angular momentum, and angular velocity of the particle in the circular orbit, respectively. Figure 3 illustrates the energy flux $F(r)$ of a disk surrounding a Schwarzschild BH embedded in a Dehnen-type DM halo with varying parameters $\rho_s$ and $r_s$. For a fixed parameter $\rho_s$, the energy flux $F(r)$ decreases as $r_s$ increases. Similarly, for a fixed parameter $r_s$, the energy flux $F(r)$ also decreases as $\rho_s$ increases. The black line represents the energy flux of a Schwarzschild (Sch) BH. For a general static spherically symmetric metric $ds^2 = g_{tt}dt^2 + g_{rr}dr^2 + g_{\theta\theta}d\theta^2 + g_{\varphi\varphi}d\varphi^2$, $E$, $L$, and $\Omega$ are denoted as

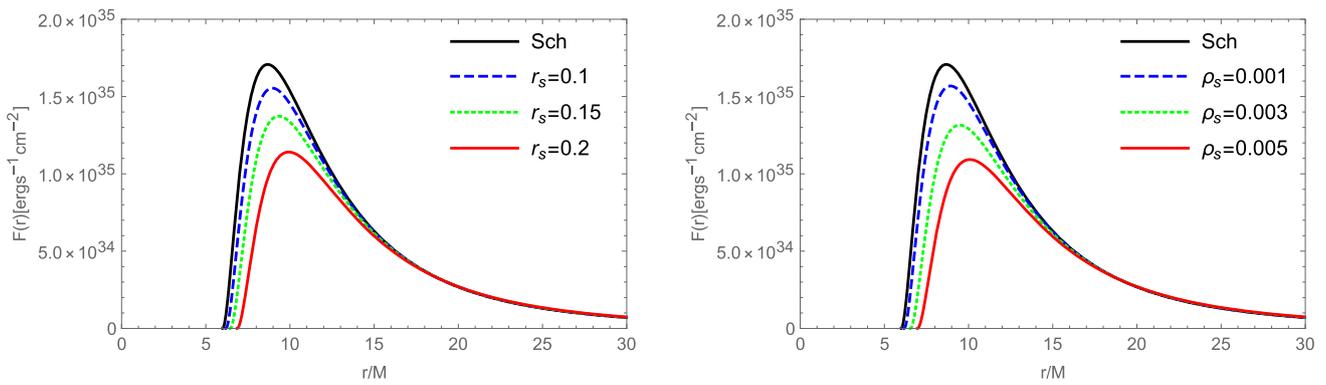

**Fig. 3** The energy flux $F(r)$ of an accretion disk around a Schwarzschild BH surrounded by a Dehnen-type DM halo for different values of the parameters $\rho_s$ and $r_s$. Left: $\rho_s = 0.03$, varying $r_s$. Right: $r_s = 0.5$, varying $\rho_s$

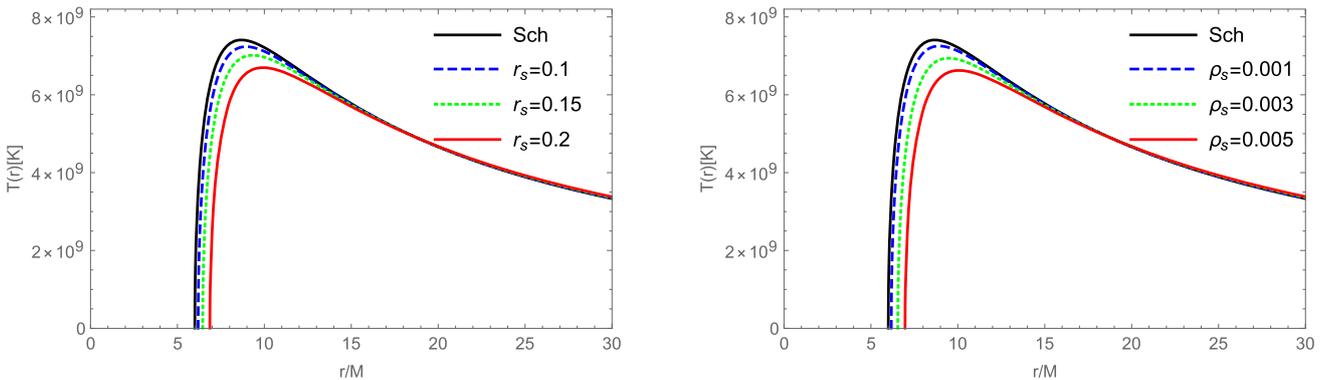

**Fig. 4** The radiation temperature $T(r)$ of an accretion disk around a Schwarzschild BH surrounded by a Dehnen-type DM halo is shown for different values of the parameters $\rho_s$ and $r_s$. Left: $\rho_s = 0.03$, varying $r_s$. Right: $r_s = 0.5$, varying $\rho_s$

$$E = -\frac{g_{tt}}{\sqrt{-g_{tt} - g_{\phi\phi}\Omega^2}}, \quad (3.2)$$

$$L = \frac{g_{\phi\phi}\Omega}{\sqrt{-g_{tt} - g_{\phi\phi}\Omega^2}}, \quad (3.3)$$

$$\Omega = \frac{d\phi}{dt} = \sqrt{-\frac{g_{tt,r}}{g_{\phi\phi,r}}}. \quad (3.4)$$

The relationship between the disk's radiation temperature $T(r)$ and its energy flux $F(r)$ is given by the Stefan–Boltzmann law: $F(r) = \sigma_{SB} T^4(r)$, where $\sigma_{SB}$ represents the Stefan–Boltzmann constant. The radiation temperature $T(r)$ of an accretion disk around a Schwarzschild BH embedded in a Dehnen-type DM halo is shown in Fig. 4. For a given value of $\rho_s$, $T(r)$ diminishes as $r_s$ grows. In a similar manner, when $r_s$ is fixed, an increase in $\rho_s$ leads to a decrease in $T(r)$.

We employ the model for the red-shifted blackbody spectrum $L(\nu)$ from [62] to calculate the observed luminosity from a thin accretion disk.

$$L(\nu) = 4\pi d^2 I(\nu)$$
$$= \frac{8\pi h \cos\vartheta}{c^2} \int_{r_i}^{r_f} \int_0^{2\pi} \frac{\nu_e^3 r}{e^{\frac{h\nu_e}{k_B T}} - 1} dr d\phi, \quad (3.5)$$

where $d$ is the distance to the disk center, $I(\nu)$ is the specific intensity at frequency $\nu$, $h$ is Planck constant, and $\vartheta$ is the disk inclination angle, assumed to be face-on ($\vartheta = 0$). The integration limits $r_i$ and $r_f$ correspond to the inner and outer radii of the disk. Following the Novikov–Thorne model, we set $r_i = r_{isco}$ and take $r_f \to \infty$, under the assumption that the local energy flux becomes negligible at large radii. The emission frequency $\nu_e$ is related to the observed frequency $\nu$ by $\nu_e = \nu(1+z)$, where the redshift factor $z$ can be expressed as [36]

$$1 + z = \frac{1 + \Omega b \sin\theta \cos\alpha}{\sqrt{-g_{tt} - g_{\phi\phi}\Omega^2}}. \quad (3.6)$$

We adopt the angle $\alpha'$ defined in Eq. (17) of Ref. [36], and relate it to the coordinate $\alpha$ used herein through $\alpha = \frac{\pi}{2} - \alpha'$. To simplify the calculation and illustrate the key physical features, we neglect light-bending effects following [63]. Under this assumption, the geometric relation simplifies to $b \cos\alpha = r \sin\phi$, allowing the redshift factor $z$ to be expressed as:

$$1 + z = \frac{1 + \Omega r \sin\theta \sin\phi}{\sqrt{-g_{tt} - g_{\phi\phi}\Omega^2}}. \quad (3.7)$$

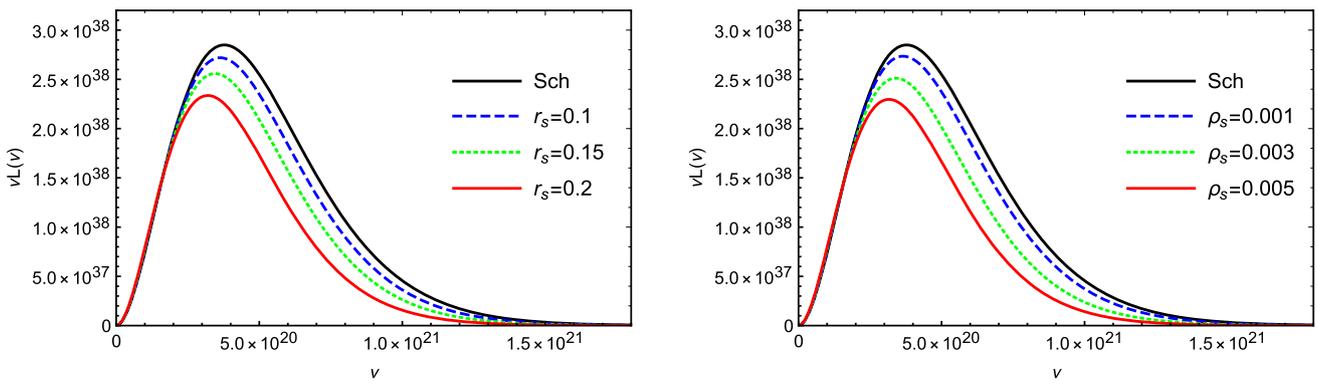

**Fig. 5** The emission spectrum $\nu L(\nu)$ of an accretion disk around a Schwarzschild BH surrounded by a Dehnen-type DM halo, shown for different values of the parameters $\rho_s$ and $r_s$. Left: $\rho_s = 0.03$, varying $r_s$. Right: $r_s = 0.5$, varying $\rho_s$

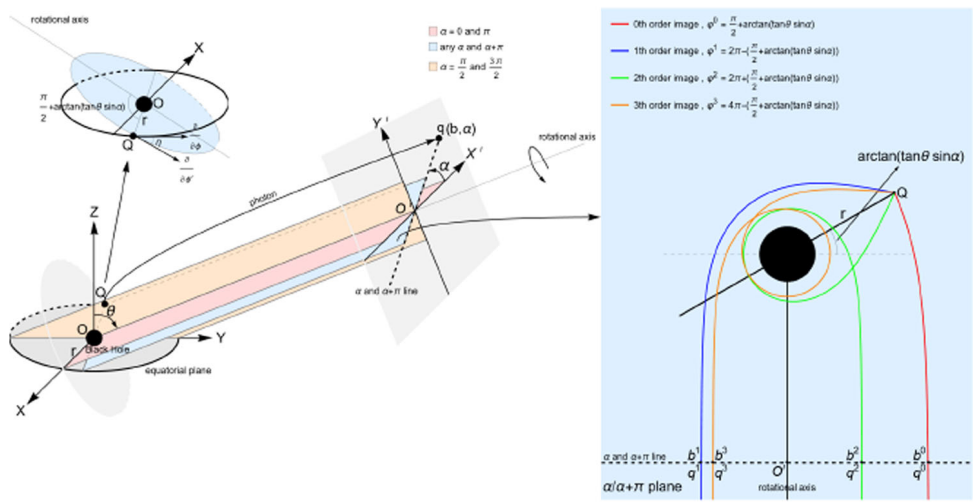

**Fig. 6** Coordinate system is indicated in Ref. [64]

Figure 5 shows the emission spectrum $\nu L(\nu)$ for an accretion disk around a Schwarzschild BH embedded in a Dehnen-type DM halo. Following the same trend as in the energy flux and radiation temperature, the presence of the DM halo suppresses the luminosity compared to the case without DM.

## 4 Images of a thin accretion disk around a Schwarzschild black hole embedded in a Dehnen-type dark matter halo

### 4.1 Observation coordinate system

The framework for our imaging analysis is the observational coordinate system depicted in Fig. 6. Here, the observer's position is fixed at $(\infty, \theta, 0)$ in the BH's $(r, \theta, \phi)$ coordinate system, with its origin at the BH's center $(r = 0)$.

To construct synthetic images, we employ backward ray tracing from the observer's image plane, parameterized by coordinates $(b, \alpha)$. Each ray, initialized at a point $q(b, \alpha)$, is integrated backward in time through the curved space-time and intersects the equatorial accretion disk at a point $Q(r, \pi/2, \phi)$. By virtue of the time-reversal symmetry of null geodesics, a photon emitted from $Q$ along the reverse path will reach the distant observer and contribute to the specific intensity at pixel $q(b, \alpha)$ in the observed image.

When maintaining a constant radial distance $r$, the resulting depiction corresponds to an orbit of a constant radius. As illustrated on the left side of Fig. 6, every $\alpha/\alpha + \pi$ plane intersects the constant-$r$ orbit within the equatorial plane at two distinct points, with their azimuthal angles $\phi$ differing by $\pi$. Specifically, in our coordinate system, the $X'$-axis is defined by setting $\alpha = 0$, while the $X$-axis aligns with $\phi = 0$. Geometrically, this configuration allows us to determine the angle $\varphi$ formed between the rotational axis and the line segment $OQ$

$$\varphi = \frac{\pi}{2} + \arctan(\tan\theta \sin\alpha). \quad (4.1)$$

As the impact parameter $b$ approaches the critical value $b_c$ associated with the photon sphere, the gravitational deflection angle diverges, enabling a single emission point $Q$ on the accretion disk to map to multiple images in the observer's

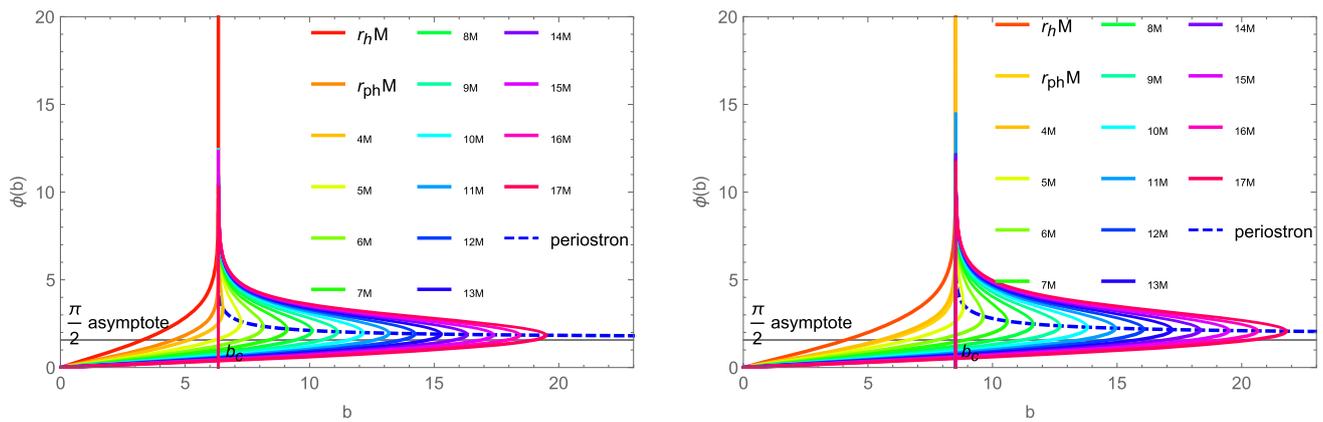

**Fig. 7** Deflection angle $\varphi(b)$ corresponding to intersections as a function of $b$ for different $r$. We set $\rho_s = 0.03$, $r_s = 0.2$ for the left side and $\rho_s = 0.03$, $r_s = 0.3$ for the right side

sky. These images, labeled $q^n$ with $n \in \mathbb{N}$, correspond to photons that wind around the BH $n$ times before reaching the distant observer, ordered by increasing azimuthal extent. Here, $n = 0$ denotes the direct image, while $n \geq 1$ represent higher-order images formed by increasingly bent null geodesics.

As shown in the right panel of Fig. 6, the relativistic images of a source point $Q$ exhibit a characteristic azimuthal pattern: even-order images ($n = 0, 2, 4, \ldots$) appear on the same side of the BH as the source, while odd-order images ($n = 1, 3, 5, \ldots$) are located on the opposite side, approximately separated by $\pi$ in azimuthal angle. The observed azimuthal coordinate of the $n$-th image is denoted by $\varphi^{(n)}$, which increases with image order due to the cumulative winding of the photon trajectory.

$$\varphi^n = \begin{cases} \frac{n}{2}2\pi + (-1)^n[\frac{\pi}{2} + \arctan(\tan\theta \sin\alpha)], & \text{when } n \text{ is even,} \\ \frac{n+1}{2}2\pi + (-1)^n[\frac{\pi}{2} + \arctan(\tan\theta \sin\alpha)], & \text{when } n \text{ is odd.} \end{cases} \quad (4.2)$$

### 4.2 Direct and secondary images of constant-$r$ orbits in the thin accretion disk

A crucial clarification is warranted regarding the "optically thick" disk model used in this work. This condition specifically refers to the disk's opacity in the vertical direction (perpendicular to the disk plane) to its own thermal emission. It implies that the radiation emitted from any point on the disk surface can be well approximated by a blackbody spectrum. However, this does not restrict the propagation paths of photons in the vacuum (or tenuous corona) outside the disk after they are emitted from the disk surface. The photons that form higher-order images (e.g., the photon ring) are emitted from the disk and then travel through the strong gravitational field of the black hole in the region outside the opaque disk material, completing multiple orbits before reaching the distant observer. Their trajectories do not pass through the interior of the optically thick disk. Therefore, the vertical optical thickness of the accretion disk is physically compatible with, and does not preclude, the formation of higher-order images via gravitational lensing in the external spacetime. Photons originating from infinity with different impact parameters $b$ intersect orbits of constant radius at distinct locations. Figure 7 illustrates the functional relationship $\varphi(b)$. As shown, for a fixed value $\rho_s = 0.03$, the curve of $\varphi(b)$ shifts rightward as $r_s$ increases. The blue dashed line in the figure, referred to as $\varphi_1(b)$, serves as a reference divider. Curves below this boundary are classified as $\varphi_2(b)$, and those above it as $\varphi_3(b)$. Accordingly, the following definitions are introduced:

$$\varphi_1(b) = \int_0^{u_{min}} \frac{1}{\sqrt{G(u)}} du, \quad (4.3)$$

$$\varphi_2(b) = \int_0^{u_r} \frac{1}{\sqrt{G(u)}} du, \quad (4.4)$$

$$\varphi_3(b) = 2\int_0^{u_{min}} \frac{1}{\sqrt{G(u)}} du - \int_0^{u_r} \frac{1}{\sqrt{G(u)}} du. \quad (4.5)$$

In the figures, each colored curve corresponds to an orbit of constant radius $r$, where any point $(b, \varphi)$ represents the deflection angle $\varphi$ of a photon with impact parameter $b$ reaching that orbit. The blue dashed line intersects each curve at its maximum, indicating the deflection angle at the photon's perihelion $r_{pe}$, the point of closest approach. This line asymptotically approaches $\varphi = \frac{\pi}{2}$, which corresponds to the limiting scenario where photons with infinitely large impact parameters $b \to \infty$ travel along straight paths tangent to the circle at $r \to \infty$ and $\varphi = \frac{\pi}{2}$.

The projection of the accretion disk onto the observer's plane is obtained by numerically solving the coupled system of Eqs. (4.1), (4.3), and (4.4), which yields the complete set of corresponding $(b, \alpha)$ pairs. Figure 8 presents the direct and secondary images of representative stable circular

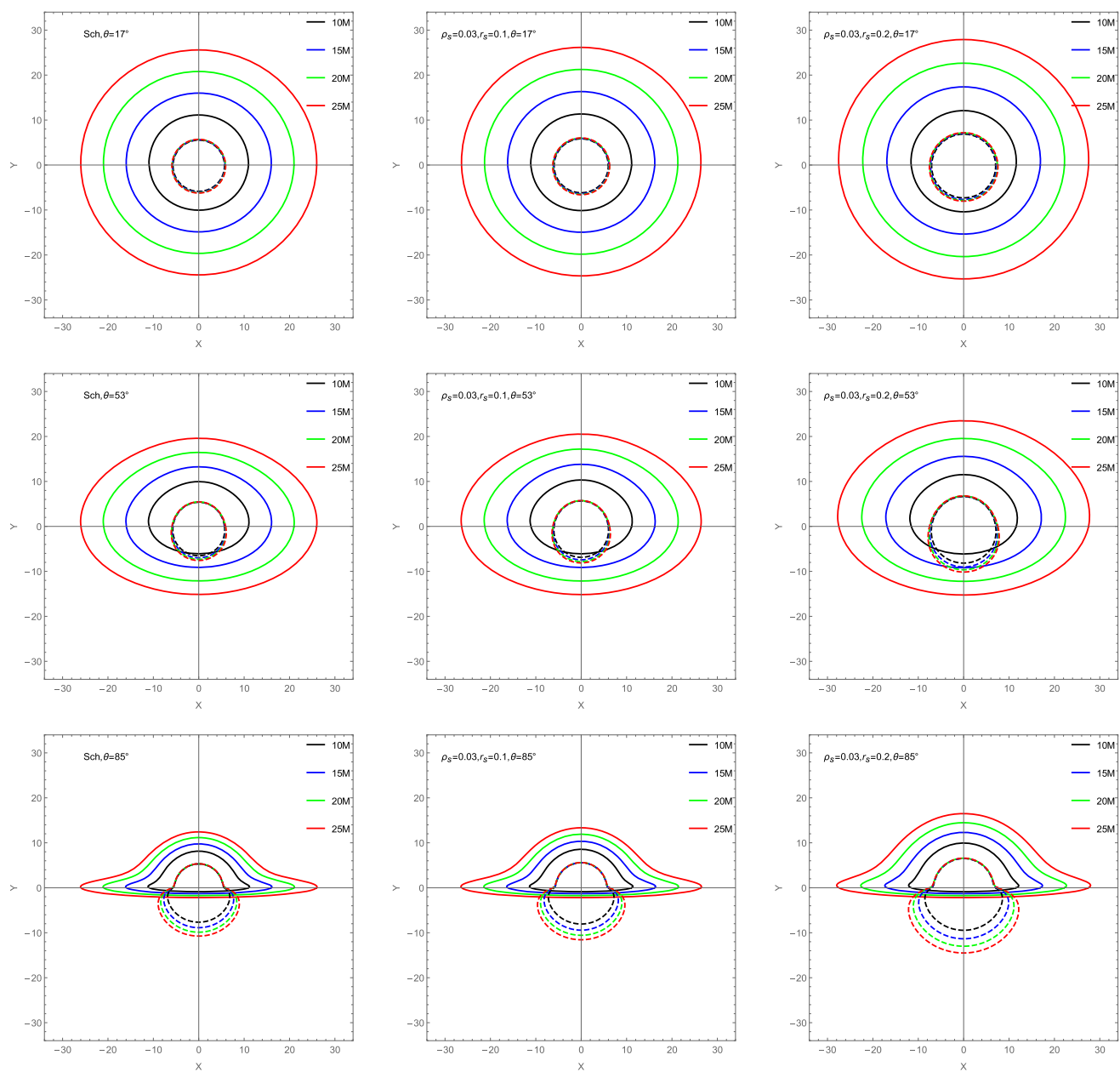

**Fig. 8** Direct (solid line) and secondary (dashed line) images of a thin accretion disk around a Schwarzschild BH embedded in a Dehnen-type DM halo, shown for different observer inclination angles. The parameter $\rho_s$ is set to 0.03, with the parameter $r_s$ being: Left: 0, Middle: 0.1, Right: 0.2

orbits around a Schwarzschild BH embedded in a Dehnen-type DM halo, as viewed by a distant observer at different inclination angles. The columns, arranged from top to bottom, correspond to inclination angles of 17°, 53°, and 85°, respectively. Each row, from left to right, corresponds to a fixed density parameter $\rho_s = 0.03$ and scale radii $r_s$ of 0, 0.1, and 0.2. These images depict stable circular orbits with radii ranging from $r = 10, 15, 20, 25$, progressing from the innermost to the outermost orbit. As the observation angle decreases, the secondary image becomes enclosed within the direct image, forming a structure reminiscent of a "photon ring", which provides a plausible alternative interpretation for such a feature. In contrast, as the observation inclination angle increases, the separation between the direct and secondary images becomes more distinct. The leftmost column represents the case of a Schwarzschild BH without DM. In comparison with the pure Schwarzschild scenario, both the direct and secondary images are found to expand outward in both horizontal and vertical directions as $r_s$ increases. Moreover, this expansion is more pronounced along the vertical direction under large inclination angles.

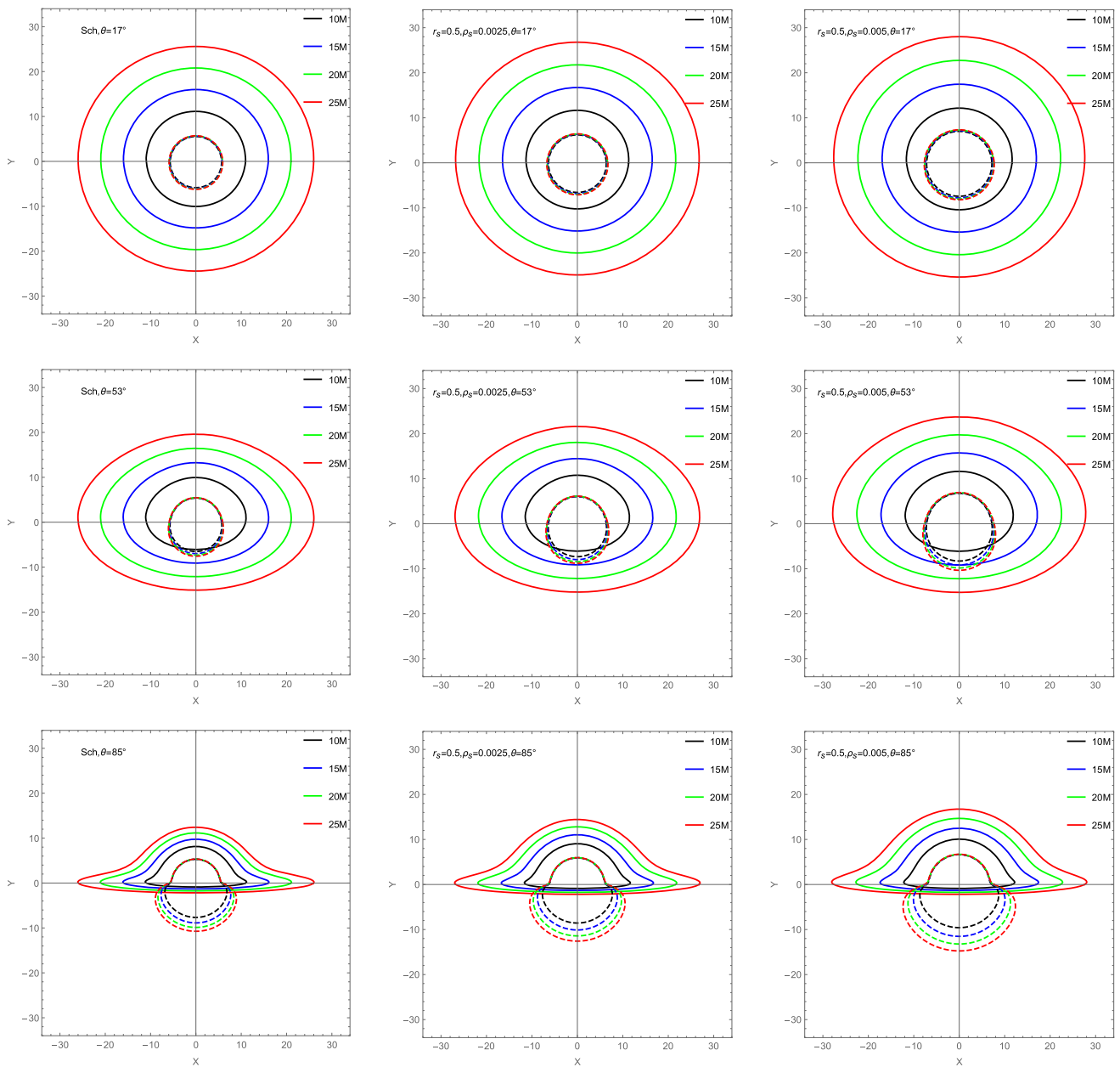

**Fig. 9** Direct (solid line) and secondary (dashed line) images of a thin accretion disk around a Schwarzschild BH embedded in a Dehnen-type DM halo, shown for different observer inclination angles. The parameter $r_s$ is set to 0.5, with the parameter $\rho_s$ being: Left: 0, Middle: 0.0025, Right: 0.005

Figure 9 displays the same configuration as Fig. 8, but with the scale radius fixed at $r_s = 0.5$ and the DM density parameter $\rho_s$ varied across columns: $\rho_s = 0$, 0.0025, and 0.005 (from left to right). The leftmost column again corresponds to the pure Schwarzschild case ($\rho_s = 0$). As $\rho_s$ increases, both the direct and secondary images expand systematically in the azimuthal and polar directions. This expansion is more pronounced at higher inclination angles, particularly along the polar (vertical) axis, indicating an increasing anisotropy in the lensing pattern.

### 4.3 Observed flux

In order to compute the observed flux $F_{obs}$ at a specific location on the image plane, the contribution of gravitational redshift $z$ must be accounted for. This leads to the relation:

$$F_{obs} = \frac{F(r)}{(1+z)^4}. \tag{4.6}$$

Combining Eqs. (3.1), (3.6), and (4.6), we obtain the observed flux as:

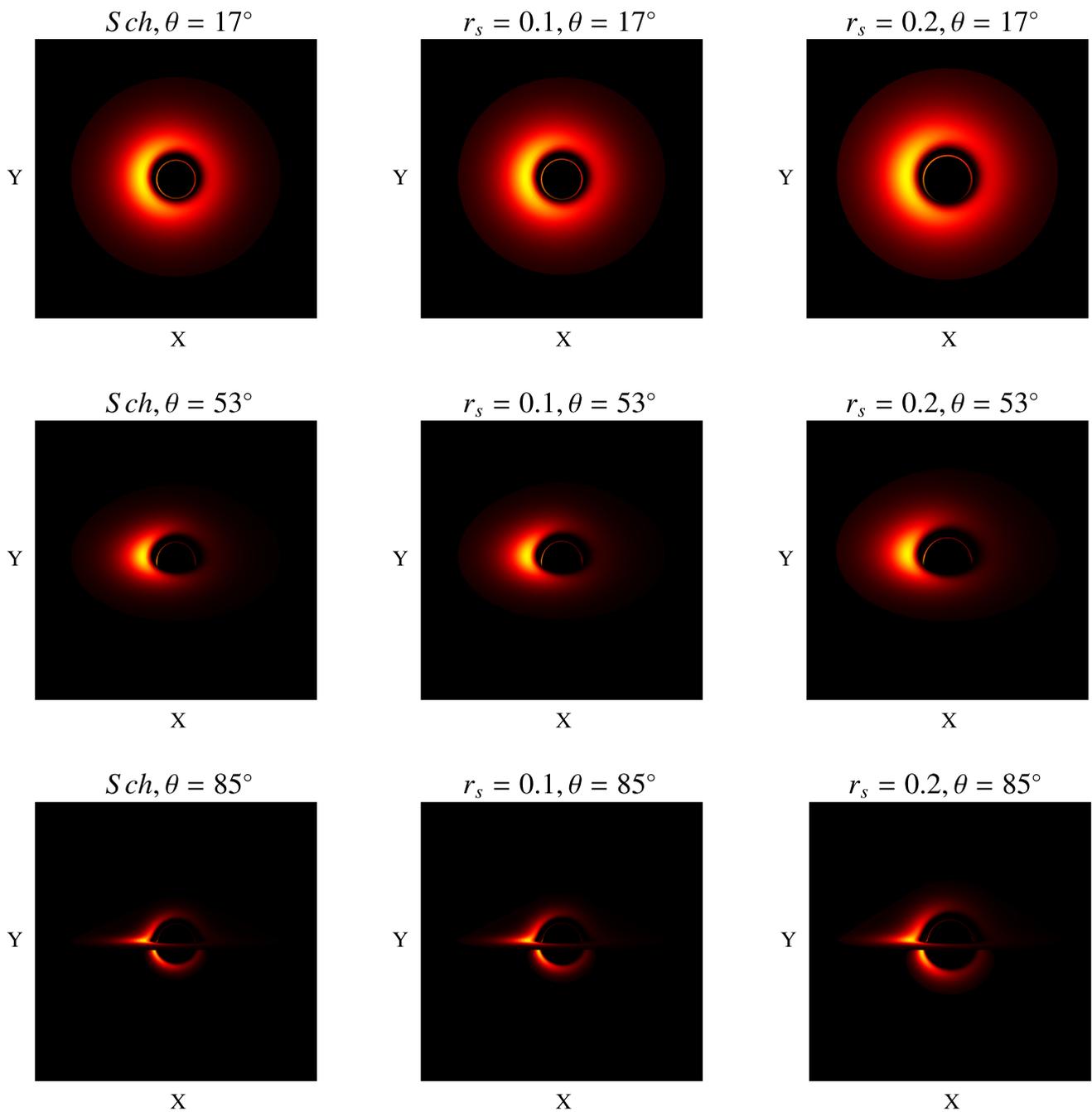

**Fig. 10** The complete apparent images of the thin accretion disk for different inclination angles. The parameter $\rho_s$ is set to 0.03, with the parameter $r_s$ being: Left: 0, Middle: 0.1, Right: 0.2

$$F_{obs} = \frac{-\frac{\dot{M}\Omega_{,r}}{4\pi\sqrt{-g}(E-\Omega L)^2}\int_{r_{isco}}^{r}(E-\Omega L)L_{,r}dr}{(\frac{1+\Omega b\sin\theta\cos\alpha}{\sqrt{-g_{tt}-g_{\phi\phi}\Omega^2}})^4}. \quad (4.7)$$

Based on the above analysis, the observed flux distribution of the accretion disk is presented in Figs. 10 and 11, which depict the imaging characteristics of a Schwarzschild BH embedded in a Dehnen-type DM halo under varying halo parameters. The morphology of the disk image is highly sensitive to the observer's inclination angle. As the inclination increases from 17° to 53° and then to 85° (top to bottom in each column), the projected image evolves from a nearly circular to a pronounced "hat-like" structure, a consequence of Doppler and gravitational beaming in the relativistic regime. In Fig. 10, rows correspond to different inclination angles, while columns (left to right) show increasing $r_s = 0, 0.1, 0.2$, with $\rho_s$ fixed at 0.03 in each row. In Fig. 11, $r_s$ is held

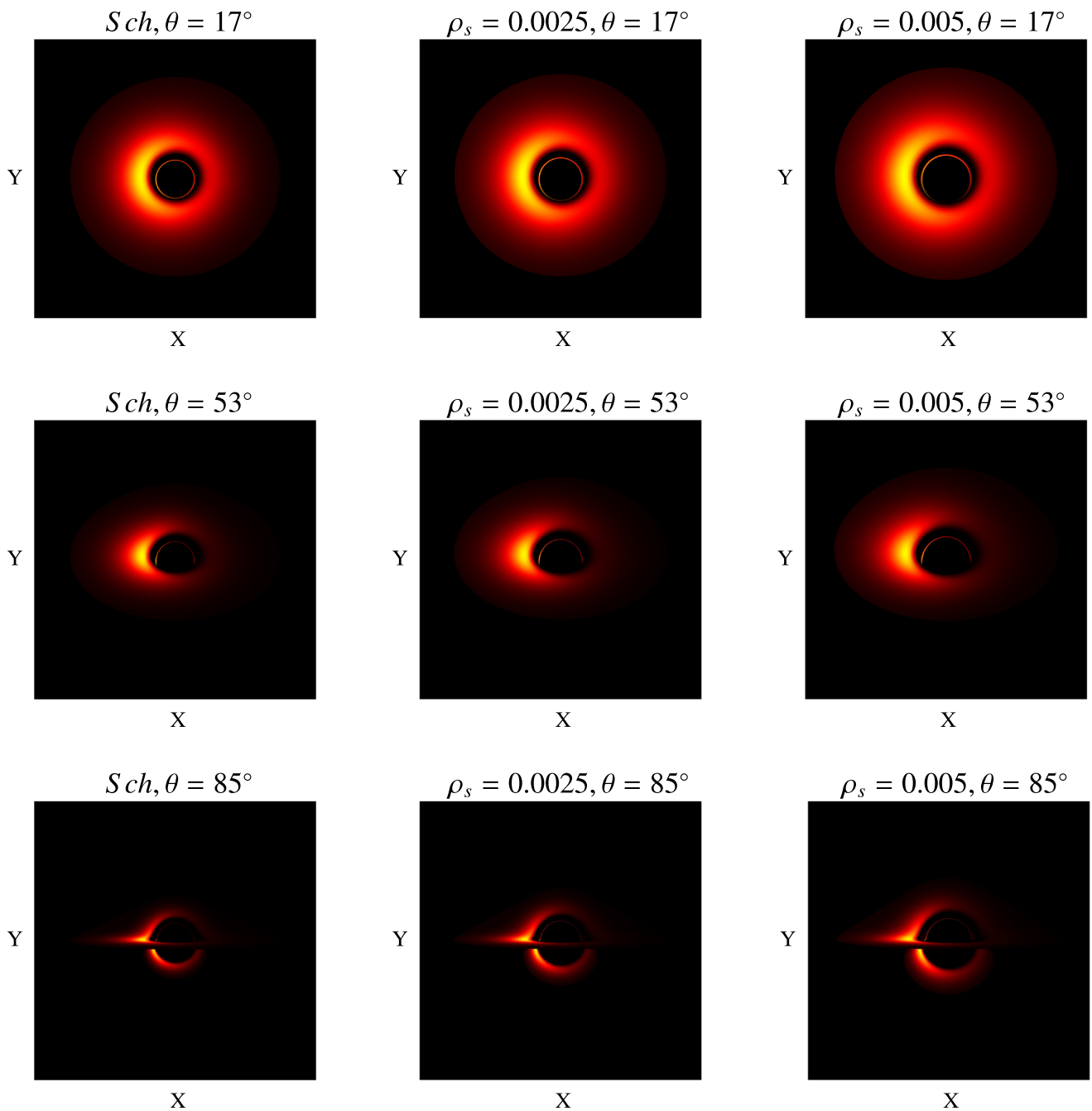

**Fig. 11** The complete apparent images of the thin accretion disk for different inclination angles. The parameter $r_s$ is set to 0.5, with the parameter $\rho_s$ being: Left: 0, Middle: 0.0025, Right: 0.005

constant at 0.5, while $\rho_s$ increases across columns: 0, 0.0025, and 0.005. The leftmost column in both figures represents the Schwarzschild case, serving as a reference. As the inclination angle increases, the flux distribution becomes increasingly asymmetric, with enhanced brightness on one side due to relativistic beaming. Despite differences in halo parameters, the overall brightness patterns across both figures exhibit quali- tative similarity, indicating a robust morphological response to viewing geometry.

## 5 Conclusion

In this work, we have investigated the physical properties and observational appearance of a geometrically thin accretion disk around a Schwarzschild BH embedded in a Dehnen-type DM halo, within the framework of the Novikov–Thorne model. Our analysis reveals that increasing the DM halo parameters-namely the density parameter $\rho_s$ and the scale radius $r_s$-systematically suppresses the energy flux, temperature distribution, and emission spectrum of the disk. This reduction occurs because the additional gravitational potential from the DM halo weakens the release of gravitational binding energy during accretion, resulting in a cooler and less luminous disk.

Moreover, $\rho_s$ and $r_s$ have a significant impact on the relativistic imaging features. As either parameter increases, both the primary (direct) and secondary images expand outward in both the azimuthal (horizontal) and polar (vertical) directions. This expansion becomes increasingly pronounced at high observer inclination angles, with a more notable elongation along the vertical axis, indicating anisotropic growth in the apparent image size.

Collectively, these results demonstrate that, under fixed conditions of BH mass, accretion rate, and viewing angle, the presence and properties of a surrounding DM halo leave distinct imprints on both the thermodynamic behavior and the visual appearance of accretion disks. Our results demonstrate that the presence of a Dehnen-type DM halo leaves distinct imprints that are, in principle, observable with instruments like the Event Horizon Telescope (EHT). Specifically, two key observational signatures could help discriminate our model from the standard Schwarzschild case: (i) A systematic suppression of the disk's radiative flux and temperature across the spectrum, for a given black hole mass and accretion rate. (ii) An increase in the apparent size of the black hole shadow and the surrounding photon ring, with a more pronounced and anisotropic enlargement at high inclination angles. Therefore, by comparing high-resolution images and spectral data from EHT with theoretical predictions encompassing DM halo effects, it may be possible to probe or constrain the properties of the central dark matter density profile in galaxies hosting supermassive black holes.

**Acknowledgements** This work was supported by the National Key R&D Program of China (grant No. 2021YFA1600400/2021YFA1600402), the National Natural Science Foundation of China (Nos. 12133011 and 12288102), and the International Centre of Supernovae, Yunnan Key Laboratory (No. 202302AN360001).

**Data Availability Statement** My manuscript has no associated data. [Author's comment: Data sharing not applicable to this article as no datasets were generated or analysed during the current study.]

**Code Availability Statement** My manuscript has no associated code/software. [Author's comment: Code/Software sharing not applicable to this article as no code/software was generated or analysed during the current study.]